\documentclass[final,5p,times,twocolumn]{elsarticle}

\usepackage{graphics}
 \usepackage{graphicx}
 \usepackage{epstopdf}
 \usepackage{epsfig}
\usepackage{amssymb}
\usepackage{amsmath}
\usepackage{lineno}

\usepackage{tikz}
\usepackage{lipsum}
\usepackage{subfig}

\begin{document}


\begin{frontmatter}



\title{Emittance Growth Due to Multiple Coulomb Scattering in a Linear Collider Based on Plasma Wakefield Acceleration}

\author{O. Mete, K. Hanahoe, G. Xia, The University of Manchester, Manchester, UK\\
The Cockcroft Institute, Sci-Tech Daresbury, Warrington, UK\\
M. Labiche, Nuclear Physics Group, STFC Daresbury Laboratory, Sci-Tech Daresbury, Warrington, UK\\
O. Karamyshev, Y. Wei, C. Welsch, The University of Liverpool, Liverpool, UK\\
The Cockcroft Institute, Sci-Tech Daresbury, Warrington, UK\\
M. Wing, University College London, London, UK, DESY, Hamburg, Germany}

\begin{abstract}
Alternative acceleration technologies are currently under development for cost-effective, robust, compact and efficient solutions. One such technology is plasma wakefield acceleration, driven by either a charged particle or laser beam. However, the potential issues must be studied in detail. In this paper, the emittance growth of the witness beam through elastic scattering from gaseous media is derived. The model is compared with the numerical studies. 
\end{abstract}

\end{frontmatter}


\section{Introduction} 
Following the success of LHC's $pp$ collisions such as the Nobel Prize winning discovery of Higgs boson, the next generation colliders for $e^+e^-$ and $ep$ collisions are vital to accomplish the tripodal scheme of discovery, precision measurements and QCD studies. Generally each successor collider should push the limits of the energy frontier further.

The concept of plasma wakefield acceleration (PWA) introduced a new technological era in high energy collider design, also proposing to employ existing infrastructure \cite{Xia}. Towards the realisation of a PWA scheme, the advantages and the issues must be explored. This study is focused on the interaction of the particles to be accelerated with the surrounding media consisting of plasma and gas. In this scope, as a first step, the growth in the emittance due to Coulomb scattering by gaseous media was studied and presented.
\section{Emittance growth due to beam-gas scattering}
\label{}     

\makeatletter{\renewcommand*{\@makefnmark}{}

Under the conditions where beam travels in the vacuum with a constant acceleration, emittance decreases with increasing energy according to the conservation of the area in the phase space given by Louville's theorem. This phenomenon is known as adiabatic damping. However, if the particles in the beam encounter a medium of gas or plasma, emittance diffusion occurs through scattering and competes against the adiabatic damping  \cite{base_theory} as suggested by the diffusion equation (Eq.\ref{eqn:diffusion_eq}); 
\begin{equation} 
\Delta\epsilon_{n,x,y}=\frac{\gamma\beta_{x,y}}{2}\overline{\mathcal{N}\langle\theta_{x,y}^{2}\rangle}
\label{eqn:diffusion_eq}
\end{equation}  
where $\gamma$ is the relativistic Lorentz constant, $\beta_{x,y}$ is the horizontal or vertical betatron function of the witness beam;  $\mathcal{N}=c n_{gas}\sigma$ is the interaction rate in the medium with $n_{gas}$ interaction centres where $\sigma$ is the scattering cross section and $c$ is the speed of light;  $\langle\theta_{x,y}^{2}\rangle$ is the mean square scattering angle. 

Beam particles can undergo elastic and inelastic scattering by the ions and electrons forming a plasma. Elastic scattering affects the particle angles and yields an emittance growth while inelastic scattering affects both particle angle and energy. Preliminarily, Coulomb scattering of the beam particles by a neutral gas media was considered to assess the impact on the beam emittance. The Coulomb scattering cross section and the mean square scattering angle are given in Eq.\ref{eqn:xsection} and Eq.\ref{eqn:theta2}:
\begin{equation}
\frac{d\sigma}{d\Omega}\approx \left( \frac{2Zr_{0}}{\gamma} \right)^{2}\frac{1}{(\theta^{2}+\theta_{min}^{2})^{2}}
\label{eqn:xsection}
\end{equation} 
\begin{equation}
\langle\theta_{x,y}^{2}\rangle = \frac{\int_0^{\theta_{max}} \! \theta^2 \frac{d\sigma}{d\Omega}  \, d\Omega}{\int_0^{\theta_{max}} \! \frac{d\sigma}{d\Omega}  \, d\Omega}
\label{eqn:theta2}
\end{equation} 
where solid angle element $d\Omega$ can be approximately given as $d\Omega \approx 2\pi \theta d\theta$ resulting in Eq.\ref{eqn:emitt1}.
\begin{equation}
\Delta\epsilon_{n,x,y}=\gamma \beta_{x,y} c n_{gas}  \int_0^{\theta_{max}} \! \pi \theta^3 \frac{d\sigma}{d\Omega} \, d\theta
\label{eqn:emitt1}
\end{equation} 
Constant acceleration along the beam axis is $\gamma(\tau)=g \tau+\gamma_{0}$ where $\gamma_{0}$ is the initial beam energy, $g$ is the rate of change of $\gamma$.
%
%
Consequently, after some calculations, emittance diffusion due to the elastic scattering of the witness beam from the neutral gas is given in Eq.\ref{eqn:emitt_diff}:
\begin{equation*}
\Delta \epsilon_{n,x,y} = \frac{2 \pi Z^2 r_0^2 c n_{gas} \beta_{x,y}}{\gamma(\tau)} \times
\end{equation*} 
\begin{equation}
\left(ln \left( \frac{\theta_{min}^2 + \theta_{max}^2}{\theta_{min}^2} \right) - \frac{\theta_{max}^2}{\theta_{min}^2+\theta_{max}^2}\right)
\label{eqn:emitt_diff}
\end{equation} 
where $Z$ is the atomic number, $r_{0}$ is the classical electron radius, $\theta$ is the scattering angle with $\theta_{min}\approx\hbar/pa$, where $a$ is the atomic radius given by $a\approx1.4\hbar^{2}/m_{e}e^{2}Z^{1/3}$, $p$ is the incident particle momentum, $\hbar$ is the reduced Planck constant, $m_e$ is the mass and $e$ is the charge of an electron. 

The total emittance at any location can be calculated as a quadratic sum of beam emittance and the contribution from the scattering process:
\begin{equation} 
\epsilon_{n,\, total}=\sqrt{\epsilon_{n}^{2}+\Delta\epsilon_{n,\, scattering}^{2}}
\label{eqn:scatt_contribution}
\end{equation}   
%
\section{Monte Carlo Simulations}
In PWA, plasma is produced by ionisation of a channel through a chamber filled with a given gas with a radius given by the ionisation laser specifications \cite{Oz}. Therefore particles travelling through the centre of the chamber may interact with the plasma ions and electrons as well as the surrounding neutral gas when they are scattered out of the plasma channel. 

A GEANT4 \cite{geant4_1,geant4_2} model was produced to study the interaction of the beam particles with plasma and gaseous medium as in the example case in fig.\ref{fig:cylinder}. As an initial step, the Coulomb scattering  of a beam of $1000$ electrons by cylindrically shaped neutral Li gas ($Z=3$, $a=6.941\,$g/mole) column with the radius of $100\,$mm was considered. A density of $6\times10^{14}$cm$^{-3}$ was chosen in order to compare the results with the previous work \cite{Allen_nature}. Li was chosen due to its orders of magnitude low scattering cross section compared to the other candidates such as Rb ($Z=37$) (see Eq.\ref{eqn:xsection}). 
\begin{figure}[htb!] 
\centering
\includegraphics[width=0.30\textwidth] {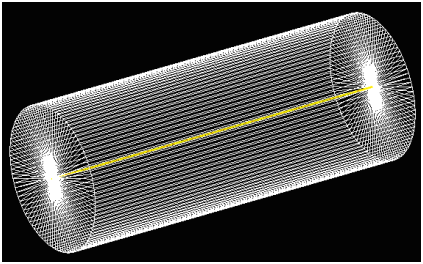}
\caption{GEANT4 simulation demonstrating the passage of a beam through Li gas. A 0.5$\,$m-long cylindrical gas column coaxial with the beam, shown in yellow, with a radius of 100$\,$mm and 1000 electrons at $10\,$GeV with $1\,$mm, $1\,$mrad initial standard deviation were considered.}
\label{fig:cylinder}
\end{figure}
The beam was tracked through the gas column with the steps of $10$ to $50\,$m. During the tracking the energy is increased linearly at each step with a gradient of 0.5$\,GeV/m$. This gradient can be provided by the plasma wakefield where the beam (the so-called witness beam) travels in the wakefields of a prior proton driver beam \cite{pwa}. Fig.\ref{fig:histos} presents example angle and position distributions after $10$ and $20\,$m travel through the plasma. It should be also noted that no means of focusing was considered for this initial simulation study including the focusing as a natural result of plasma wakefields. Effects of the various focusing schemes are under study and will be presented elsewhere.
\begin{figure}[htb!] 
\centering
\includegraphics[width=0.52\textwidth] {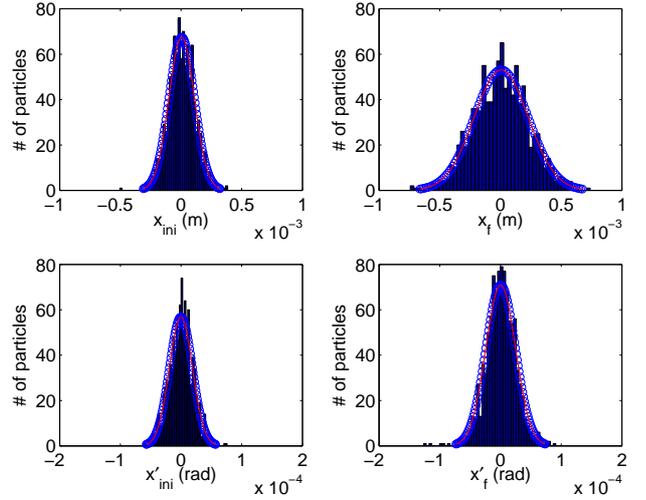} 
\caption{An example for the position (top) and angle (bottom) distributions of the beam, after $10$ (left) and $20\,$m (right) travel through Li gas.}
\label{fig:histos}
\end{figure} 
The initial beam distribution in the phase space, resulting in a $1\,\pi\,$nm$\,$rad is given in fig.\ref{fig:phase_space}-a, was implemented in GEANT4 primary particle generation routine. Only Coulomb scattering was activated in the physics list for comparison with the theory where only Coulomb scattering cross-section is considered. The resulting phase space evolution in the entrance and exit of various steps are shown in fig.\ref{fig:phase_space}-b, c and d, respectively. 
\begin{equation}
\epsilon = \sqrt{\langle x^2 \rangle \langle x'^2 \rangle - \langle xx' \rangle^2 }
\label{eqn:emitt}
\end{equation} 
For the purposes of assessing a long PWA acceleration section, beam was tracked up to 500$\,$m within Li gas. Phase space was reconstructed at various positions and the resulting emittance is calculated according to Eq.\ref{eqn:emitt} and presented in fig.\ref{fig:emitt_results} in comparison with theoretical curves based on Eq. \ref{eqn:emitt_diff}. The two theoretical curves follow the trend in simulated data. The red curve includes the theoretical emittance growth normalised by a factor to be able to fit the data. The reason for the discrepancy is under study. 
\section{Conclusions and Outlook}
As an advanced accelerating technique PWA has ever-increasing prospects. Therefore, the potential issues of the scheme must be assessed carefully. This study was initiated to seek out the impact of interaction of a witness beam with the surrounding plasma formed to provide acceleration. The emittance growth induced in the beam via beam-gas elastic scattering was studied analytically based on the existing model for the beam-gas scattering in damping rings. 
\begin{figure}[htpb!]  
\centering
\subfloat[]{\includegraphics[width=0.36\textwidth] {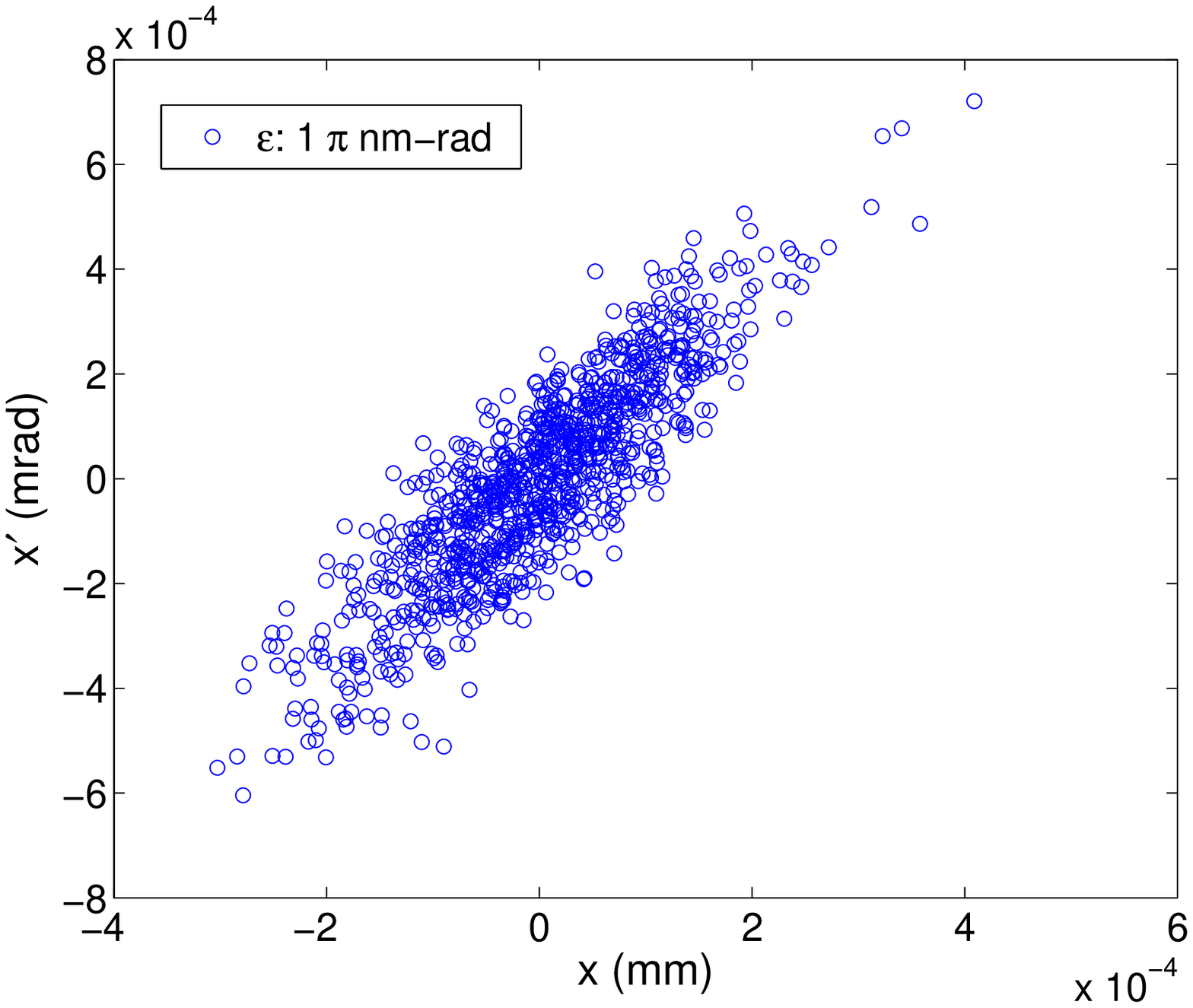}} \\
\subfloat[]{\includegraphics[width=0.36\textwidth] {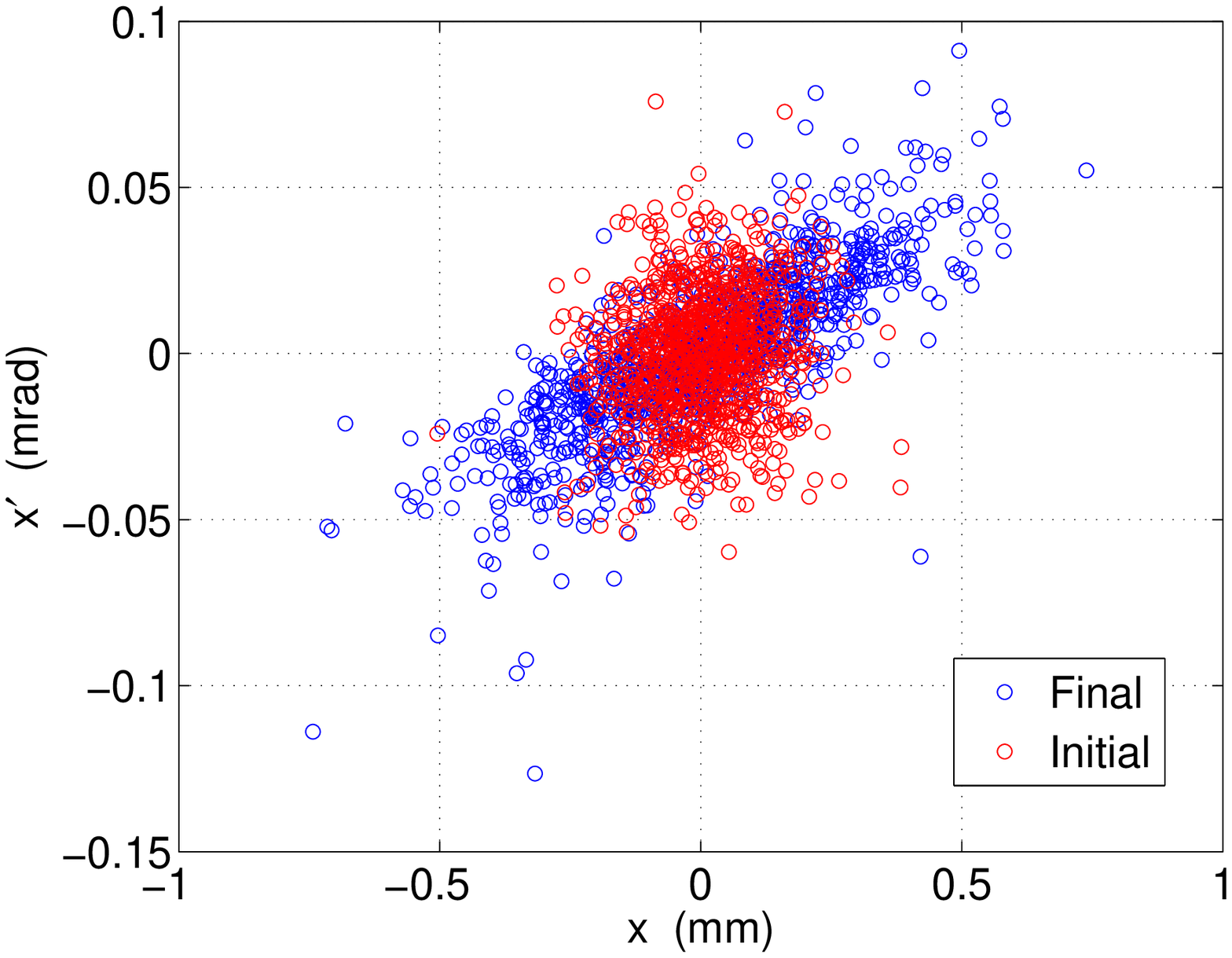}} \\
\subfloat[]{\includegraphics[width=0.34\textwidth] {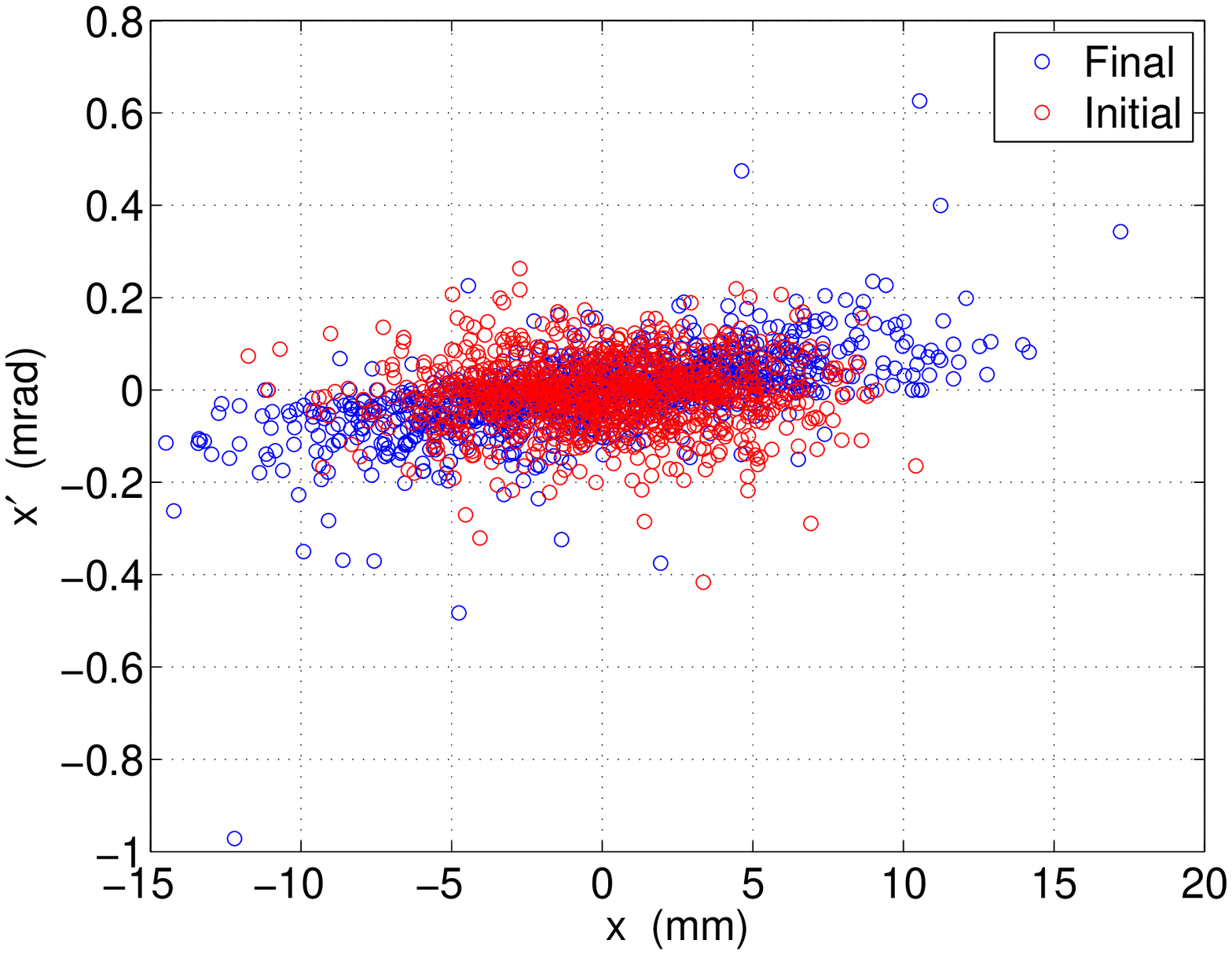}} \\
\subfloat[]{\includegraphics[width=0.34\textwidth] {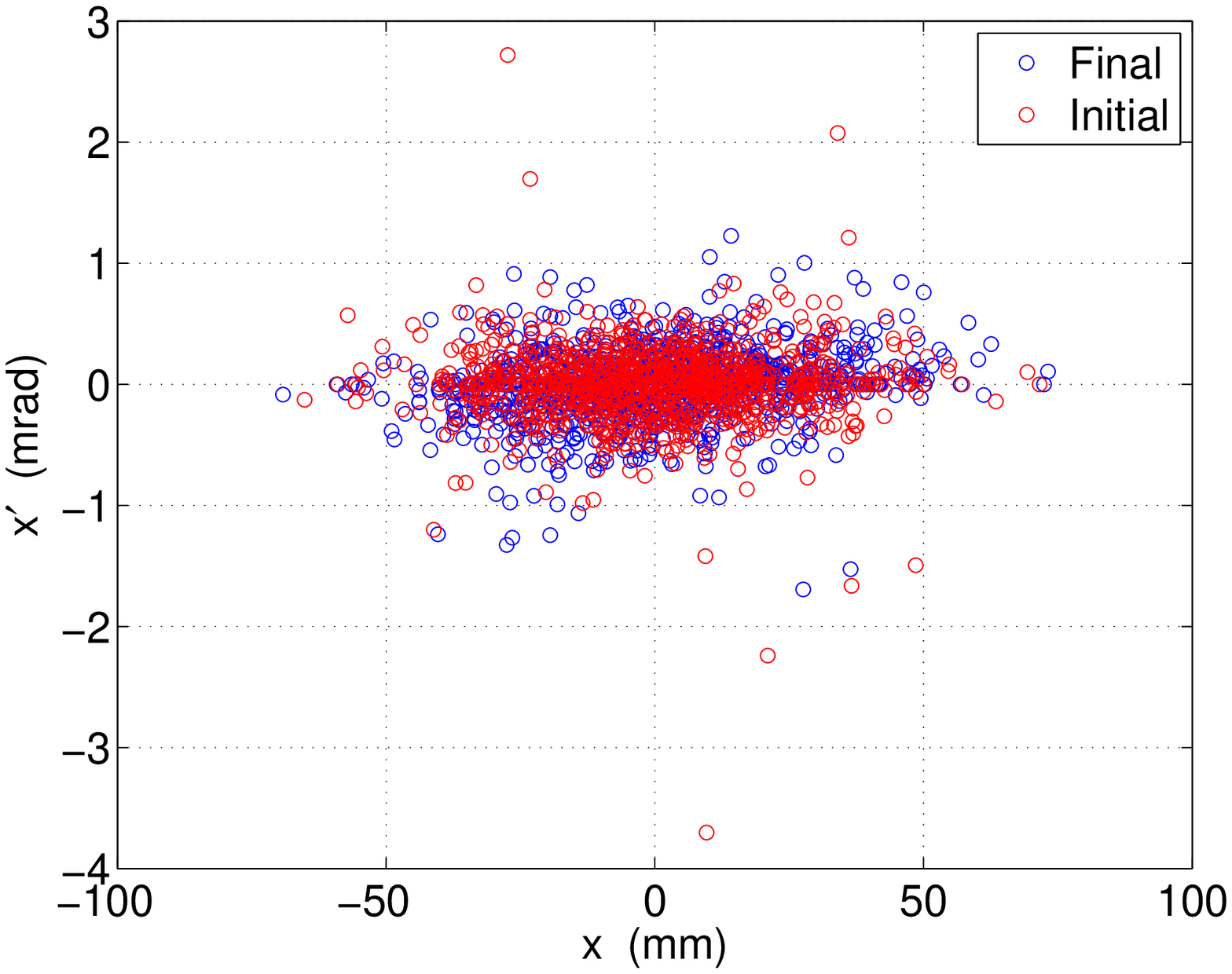}} \\
\caption{a) Initial phase space distribution and evolution of phase space for the electron beam travelling through 500$\,$m neutral Li gas sampled and compared at b) 10-20$\,$m, c) 130-180$\,$m, d) 480-500$\,$m as initial and final, respectively.}
\label{fig:phase_space}
\end{figure}
\begin{figure}[htb!] 
\centering
\includegraphics[width=0.436\textwidth] {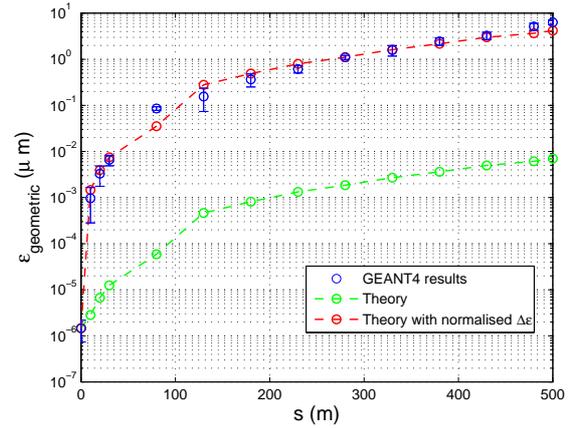}
\caption{Evolution of the emittance along 500$\,$m long Li gas while undergoing Coloumb scattering in comparison with theoretical curves.}
\label{fig:emitt_results}
\end{figure}

The preliminary results of this novel study are presented in this paper. It has been shown that the beam-plasma interaction is expected to be significant in the absence of any means of focusing and the evolution of phase space can be feasibly studied by using GEANT4. In order to obtain the most realistic representation of interaction between beam and plasma and a realistic estimation on the emittance growth, several improvements to the existing GEANT4 model are in progress. Consequently, in the further studies the model will include an electric field on the beam axis to simulate the linear acceleration instead of changing the beam energy at each step of simulation. In addition a magnetic field will be introduced in a way to simulate the focusing component of the plasma wakefield. A thin plasma channel consisting of ionised gas and electrons will be simulated within a volume of gas atoms to represent the configuration in an actual plasma tank where ionisation occurs only in the inner region illuminated by a laser. 
\section{Acknowledgements}
The authors would like to thank Dr. Robert Apsimon (CERN), Dr. Jakob Esberg (CERN) and Dr. Frank Simon (Max Planck Institute) for very useful discussions on beam tracking and phase space analysis.  

This work was supported by the Cockcroft Institute Core Grant and STFC.
%

\end{document}